\documentclass[twocolumn,preprintnumbers,aps,amsmath,amssymb,prb]{revtex4}

\usepackage{graphicx}
\usepackage{dcolumn}
\usepackage{bm}

\begin{document}


\title{Liquid-liquid phase transitions in supercooled water studied by computer simulations of various water models}
\author{Ivan Brovchenko}
\email{brov@heineken.chemie.uni-dortmund.de}
\author{Alfons Geiger}
\email{alfons.geiger@udo.edu}
\author{Alla Oleinikova}
\email{alla@heineken.chemie.uni-dortmund.de}
\affiliation{Physical Chemistry, Dortmund University, 44221 Dortmund, Germany}
\date{\today}
\begin{abstract}
Liquid-liquid and liquid-vapor coexistence regions of various water models were determined by  MC simulations of isotherms of density fluctuation restricted systems and by Gibbs ensemble MC simulations. All studied water models show multiple liquid-liquid phase transitions in the supercooled region: we observe two transitions of the TIP4P, TIP5P and SPCE model and three transitions of the ST2 model. The location of these phase transitions with respect to the liquid-vapor coexistence curve and the glass temperature is highly sensitive to the water model and its implementation. We suggest, that the apparent thermodynamic singularity of real liquid water in the supercooled region at about 228 K is caused by an approach to the spinodal of the first (lowest density) liquid-liquid phase transition. The well known density maximum of liquid water at 277 K is related to the second liquid-liquid phase transition, which is located at positive pressures with a critical point close to the maximum. A possible order parameter and the universality class of liquid-liquid phase transitions in one-component fluids is discussed. \end{abstract}
\maketitle
\section{\label{sec:Intro}Introduction}
Some one-component substances show several amorphous states at low temperatures. This suggests the possible existence of phase transitions between them. Evidence for such phase transitions in supercooled water was obtained both from experiments \cite{Mishima85,MishimaSt,Mishima02,Klotz} and simulations \cite{Poole92,Poole93a,Har97a,Sciort02,Sciort03,BGO03,Kyoto,Jedl} (see also recent reviews \cite{DebSt,Debrev,Angellrev,Stanleyrev}). The presence of phase transitions between different liquid (glassy) states of an isotropic fluid in the supercooled state (below the freezing temperature) may noticeably influence its properties above the freezing temperature. In particular, at low temperatures an anomalous, singularity-like behavior of some thermodynamic properties is observed in several fluids, \cite{Voronel} including water. \cite{Anisimov1,heat1old,heat1} Such behavior can be explained by the influence of a distant liquid-liquid phase transition with a critical point located in the supercooled region.\cite{Poole92}  
\par Liquid-liquid phase transitions of a one-component fluid may occur in a thermodynamic region where one or both liquid phases are metastable with respect to crystallization or evaporation. This makes experimental studies of such phase transitions difficult or even impossible. Fortunately, this is not the case for computer simulations, which can be used to study fluids in metastable or even unstable states. In principle, coexistence of two liquid phases with an explicit interface between them can be obtained by direct simulations in the constant-volume \textit{NVT} ensemble. However, this demands the use of large systems, which can not be equilibrated at low temperatures during reasonable computer simulation times. Decreasing the system size causes a narrowing of the density range, where the coexistence of two phases with explicit interface can be obtained. \cite{Binder} In a small enough system, whose size does not exceed some "threshold" value, a phase separation with a persistent interface can be prevented completely and a continuous isotherm, joining the two states at subcritical temperatures, can be simulated. The presence of a van der Waals-like loop directly evidences the sub-critical character of an isotherm. This allowed to detect a liquid-liquid phase transition in ST2 \cite{Poole92,Poole93a,Poole93b,Poole97a,Poole97b,Sciort03} and TIP5P \cite{Sciort02} water models. Note, that the unavoidable tendency of the simulated metastable and unstable systems to phase separation ultimately distorts the computed isotherms. \cite{Hansen} Moreover, the threshold value of the system size, which prevents phase separation in the simulation box in the whole density range of the two-phase coexistence region, can not be estimated in advance. These factors represent intrinsic limitations of the applicability of simulations in the \textit{NVT} ensemble to study phase transitions, which can not be overcome by varying the system size or by improving the sampling. In particular, these shortcomings can lead to the absence of a van der Waals-like loop in a sub-critical isotherm close to the critical temperature, resulting in a lower value of the estimated critical temperature of the phase transition.  
\par         
Simulations in the constant-pressure \textit{NPT} ensemble can also be used to study low-temperature liquid-liquid phase transitions. An abrupt change of the density along an isobar at some temperature indicates a phase transition. However, it is difficult to distinguish a phase transition from a sharp but continuous change of the density. \cite{Tanaka96,Paschek} Only stable and metastable but not mechanically unstable states can be explored in the \textit{NPT} ensemble and, therefore, a continuous isotherm, joining two phase states at subcritical temperatures, can not be obtained. Simulations of subcritical isotherms in the \textit{NPT} ensemble are accompanied by hysteresis phenomena, which may serve as an indicator of a  first-order phase transition. However, the extension of a hysteresis loop depends on the system size and simulation time. As in the case of the \textit{NVT }ensemble, the extension of such simulations beyond the stable states could hardly be controlled.         
\par 
The study of phase transitions by simulations in canonical ensembles can be substantially improved by forcing the system to remain homogeneous even in metastable and unstable states. This can be achieved by restricting the density fluctuations on a mesoscopic scale. \cite{Hansen,Debrest} Simulations of isotherms in such restricted ensembles enable the location of liquid-liquid phase transitions at supercooled temperatures. \cite{BGO03,Kyoto} In the present paper we use this method to find liquid-liquid phase transitions of various water models in the supercooled region. In some cases the densities of the coexisting liquid phases were determined directly by simulations in the Gibbs ensemble. \cite{BGO03,Kyoto} To locate liquid-liquid transitions with respect to the liquid-vapor coexistence curve, the latter was simulated in the Gibbs ensemble down to low temperatures. The obtained phase diagrams are used to analyze the sensitivity of the liquid-liquid and liquid-vapor phase transition to the peculiarities of a given water model and its implementation. From these data the possible location of the liquid-liquid phase transitions in real water is inferred.   
\section{\label{sec:Simmethods} Simulation Methods}
ST2, \cite{ST2} TIP4P, \cite{TIP4P} TIP5P \cite{TIP5P} and SPCE \cite{SPCE} water models were used in our studies. In all cases a simple spherical cutoff \textit{R$_c$} of the intermolecular interactions was used with long-range corrections for the Lennard-Jones potentials (LRLJ corrections). The value of \textit{R$_c$} was 9 $\mbox{\AA}$ in all cases, except the TIP4P model, where \textit{R$_c$} was 8.5 $\mbox{\AA}$ in the simulations of the liquid-vapor coexistence curve. In accord with the original parametrization of these models, \cite{ST2,TIP4P,TIP5P,SPCE} no long-range corrections for the Coulombic interaction (LRCI corrections) were included. Additionally, we used the ST2 water model including a reaction field to account for LRCI corrections (ST2RF water model hereafter), which has already been studied intensively in the supercooled region. \cite{Poole92,Poole93a,Poole93b,Poole97a,Poole97b,Sciort03} The liquid-vapor coexistence curve of various water models was obtained by Monte Carlo (MC) simulations in the Gibbs ensemble (GEMC).\cite{GE} The coexistence of different water phases at supercooled temperatures was studied by GEMC simulations and by simulations of isotherms in the density fluctuation restricted \textit{NPT} ensemble.\cite{Hansen,Debrest}
\subsection{\label{sec:GE} Gibbs ensemble MC simulations of the liquid-vapor and liquid-liquid coexistence}
The liquid-vapor coexistence of water was simulated by direct equilibration of the two coexisting phases in the Gibbs ensemble.\cite{GE} The total number \textit{N} of molecules in the two simulation boxes was about 600, except for the case of the TIP4P model, where \textit{N} was about 300 to 400, depending on temperature. The use of efficient techniques for the molecular transfers (more details can be found elsewhere \cite{BGO2004}) allowed to extend the simulated coexistence curves deeply into the supercooled region. Even at extremely low temperatures the number of successful molecular transfers in the course of the simulation runs always exceeded 10\textit{N} and achieved 100\textit{N} at high temperatures. The obtained liquid densities did not depend on the initial configuration of the simulated system even at the lowest studied temperatures. This was achieved by efficient sampling of the system due to molecular transfers and also by the absence of phase separation in the simulation box, which is intrinsically avoided in the Gibbs ensemble. The critical temperature and critical density of the liquid-vapor coexistence curves were estimated by fitting the order parameter with a scaling law including one correction term and by fitting the diameter to a second order polynomial. The coexisting densities above the temperature of the liquid density maximum were used for this fitting procedure.
\par
The coexistence between liquid water phases of different densities was also studied by GEMC simulations. The extremely low acceptance probability for molecular transfers between two dense liquid phases practically prevents such simulations at temperatures below $\sim$260 K and densities above $\sim$1.3 g cm$^{-3}$. Another complication is caused by the possible small difference between the densities of the coexisting liquid phases. GEMC simulations develop two coexisting phases, when the average density $\rho_{av}$ of the two simulation boxes is in the two-phase region. The initial choice of this average density determines the box sizes and numbers of molecules in both boxes, which have to be adequate to reproduce the properties of both coexisting phases. This essentially restricts the range of $\rho_{av}$ , which are appropriate for such studies. Therefore, we have equilibrated several initial configurations of liquid water with different densities in the interval from 0.90 to 1.25 g cm$^{-3}$ by MC simulations in the \textit{NVT} ensemble. Then, several pairs of these configurations with density differences from 0.10 to 0.20 g cm$^{-3}$ were used as starting configurations for GEMC simulations. The number of successful molecular transfers was about \textit{N} at \textit{T} = 260 K, about 2\textit{N} at \textit{T} = 270 K and essentially more at higher temperatures. Any choice of the average density $\rho_{av}$ at \textit{T} $>$ 270 K resulted in comparatively fast (less one month of computer time on a GHz processor) equalization of the densities in the two boxes. The same result was obtained also at \textit{T} = 260 or 270 K, when $\rho_{av}$ $\geq$ 1.10 g cm$^{-3}$. However, when the average density $\rho_{av}$ was about $\sim$0.95 and $\sim$1.05 g cm$^{-3}$ the water densities in the two boxes did not tend to equalize at \textit{T} = 260 and 270 K even during extremely long simulation runs (several months of computer time on a GHz processor) but a stable liquid-liquid equilibrium was reached. It should also be noticed, that we have failed to get the coexistence of two liquid water phases, when starting from two equal densities in the two simulation boxes.  
\begin{figure}[t]
\includegraphics[width=6cm]{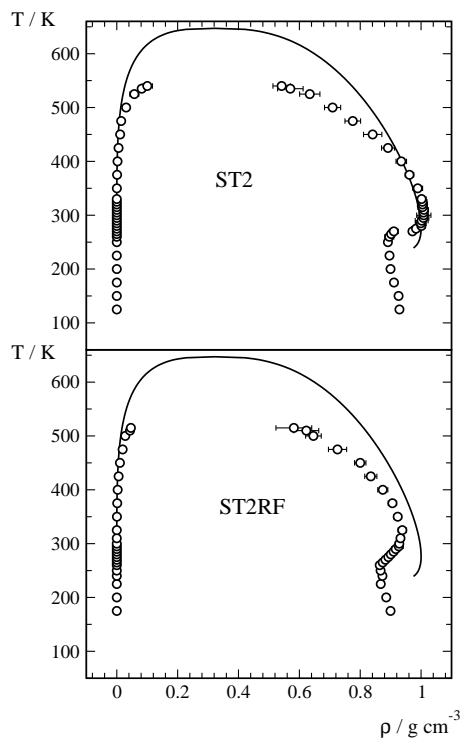}
\caption{ Liquid-vapor coexistence curve of the ST2 and ST2RF water models (circles). Experimental data \cite{CCexp,CCexpZ} are shown by the solid lines.}
\end{figure}      
\subsection{\label{sec:NPT}MC simulations in the restricted NPT ensemble}
An extension of the isotherm into metastable or unstable regions can be achieved by simulations in the density fluctuation restricted ensemble,\cite{Hansen,Debrest} which allows the simulation of an artificial single-phase state. The simulation box with \textit{N} molecules has to be divided into \textit{n} equal subcells. In the general case, the number \textit{N$_i$} of molecules in the \textit{i}-th subcell should satisfy the following constraint:
\begin{eqnarray}
\label{eq:deb}
\textit{N/n} - \delta \textit{N} \leq N_i \leq \textit{N/n} + \delta \textit{N},
\end{eqnarray}
where $\delta$\textit{N} is the maximum deviation of the number of molecules in the subcell from the average value \textit{N/n}, allowed in simulations. Such a constraint can be fulfilled in MC simulations by rejecting those moves, which violate Eq.\ref{eq:deb}. The value \textit{N/n} should be small enough to exclude phase separation in each subcell, whereas too small values of \textit{N/n} could result in the freezing of the simulated liquid. MC simulations in the restricted \textit{NVT} ensemble were successfully applied for reproducing the liquid-vapor coexistence curve of a LJ fluid. \cite{Hansen,Debrest} In the case of liquid-liquid coexistence in a one-component fluid the densities of the coexisting phases could have rather close values (difference $<$ 10 $\%$). So, even $\delta \textit{N}$ = 1 could result in a phase separation in a simulation box with several hundreds molecules. Therefore, we imposed $\delta$\textit{N} = 0, which could noticeably suppress the density fluctuations in the \textit{NVT} ensemble. \cite{Hansen,Debrest} Imposing this local restriction in an \textit{NPT} simulation, the fluctuation of the average density  is not restricted. This method can be applied for simulations of stable and metastable, but not for unstable states.
 \begin{figure}[t]
\includegraphics[width=7cm]{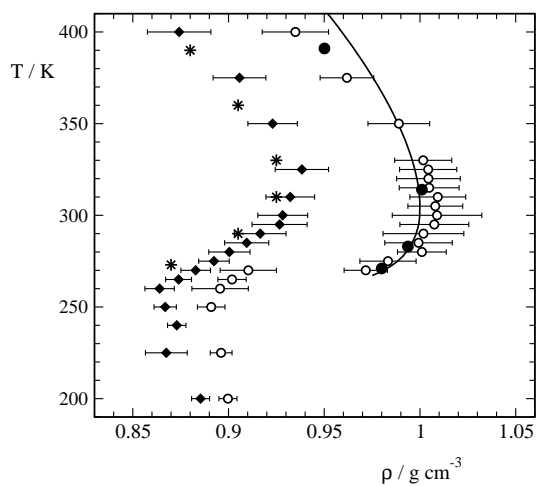}
\caption{ Liquid densities of ST2 (open circles) and ST2RF (solid diamonds) water models along the liquid-vapor coexistence curve. Simulation data at \textit{\textit{P}} $\approx$ 0 for the ST2 model \cite{ST2} and the ST2RF model \cite{Poole93a} are shown by solid circles and stars, respectively. Experimental data,\cite{CCexp,CCexpZ} shifted upwards by 28$^{\circ}$, are shown by the solid line.}
\end{figure}     
\par
In our implementation of the fluctuation restricted \textit{NPT} ensemble, the cubic simulation box with 513 water molecules was divided into 27 equally sized cubic subcells which contain an equal number (19) of molecules. The number of molecules in each subcell was kept unchanged in the course of the MC simulations. Typically up to 2*10$^5$ molecular moves per molecule were done in the course of a MC simulation run. Several isotherms between 150 and 300 K were simulated for each studied water model. To test the effect of the restrictions on the thermodynamic properties of the model water, we simulated the isobar \textit{P} = 0 in the restricted \textit{NPT} ensemble for the ST2 model from 200 to 450 K. The liquid densities obtained by this method were compared with the saturated liquid densities, obtained by the GEMC simulations (see Fig.2). Additionally, the isotherm \textit{T} = 260 K of the ST2 model was simulated in a simulation box with 621 molecules, divided into equally sized cubic subcells which contain 23 molecules each.  
\begin{figure}[t]
\includegraphics[width=6cm]{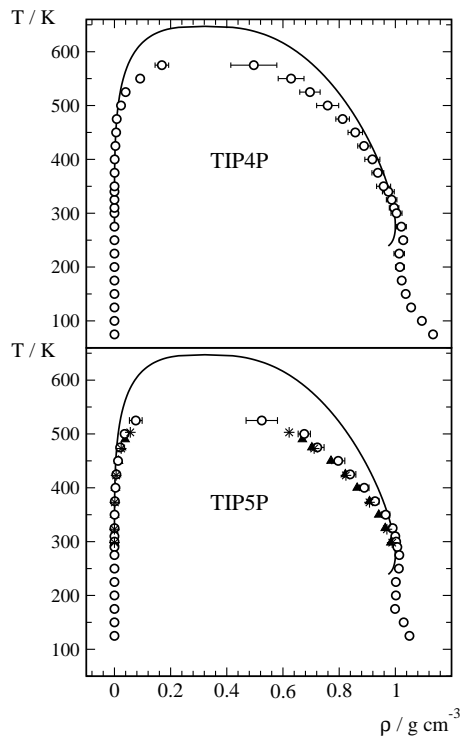}
\caption{ Liquid-vapor coexistence curve of TIP4P and TIP5P water models (circles). Simulation data of Ref.\onlinecite{Nez02} and Ref.\onlinecite{Zeng02} are shown by triangles and stars, respectively. Experimental data \cite{CCexp,CCexpZ} are shown by solid lines.}
\end{figure} 
\par
Since neither the location of the liquid-liquid phase transitions (except for one transition of the ST2RF model \cite{Poole92,Poole93a,Poole93b,Poole97a,Poole97b,Sciort03}) nor even their number are known a priori, we explored in detail a wide interval of density (from $\sim$0.8 to $\sim$1.4 g cm$^{-3}$) and pressure (from $\sim$-6 to $\sim$+10 kbar). To detect all branches of the studied isotherm, we performed simulations in the following way. Several configurations of liquid water with different densities were equilibrated in the restricted \textit{NVT} ensemble. Then, each prepared configuration was used as an initial one for simulations in the restricted \textit{NPT} ensemble at several chosen pressures. The systems converge to one (at supercritical temperatures) or more (at subcritical temperatures) densities, which correspond to all possible stable and metastable states of liquid water at the considered pressure and temperature. At low temperatures such convergence demanded extremely long simulation runs (several months in some cases). Subsequently, the density along the isotherms in each detected water phase was studied by increasing (decreasing) the pressure in small steps (0.1 kbar in some cases). At some pressure, which roughly corresponds to the stability limit (spinodal), the density of a considered water phase in the course of a simulation run rapidly shifts to another branch of the isotherm, corresponding to another water phase. This approach should allow for discovering all branches of a water isotherm, whose densities at the same pressure differ more than about 0.03 g cm$^{-3}$. This value approximately corresponds to the thermodynamical fluctuations of the density in our finite system. Note, that we checked the influence of the initial conditions of the simulation runs. Even at the lowest temperatures we observed convergence to the given results. This was due to the extremely long simulation runs and the absence of phase separation in the simulation box, which is intrinsically avoided in the restricted ensemble. 

\par
The obtained isotherms do not allow to locate accurately the coexistence pressure and the densities of the coexisting phases due to the missing parts in the region of the unstable states. Moreover, the density is not necessarily the appropriate order parameter to describe the liquid-liquid phase transition of a one-component system. Therefore, we attributed the coexistence pressure to the average of the maximal and minimal pressures of the overlap region, where the two branches have two distinct densities at the same pressure. Above the glass transition temperature \textit{T$_g$}, the accuracy of such estimations is comparable with the accuracy of the GEMC simulation data. Far below \textit{T$_g$}, the \textit{density} ranges of the different branches of the isotherm overlap so strongly, that the densities of the coexisting phases can be estimated only very roughly.  Note, that at \textit{T} $<$ \textit{T$_g$}, we do not infer the existence of some phase transition from abrupt variations of the density in a small pressure interval. Additional information about the sign of the coexistence pressure we can obtain by attributing the density of the liquid branch of the liquid-vapor coexistence curve at the same temperature to zero pressure.      
\begin{table}[t]
\caption{The values of the liquid-vapor critical temperature \textit{T$_c$} and critical density $\rho_c$, temperature of the liquid density maximum \textit{T$_{max}$} at the liquid-vapor coexistence curve or at zero pressure, temperature of the glass transition \textit{T$_g$}, obtained from simulations of the studied water models. Experimental data are also shown.}
\label{tab:par} 
\vspace{0.5cm}     
\begin{tabular}{l|c|c|c|c|c}
\hline\noalign{\smallskip}
Model &  \textit{T$_c$}  & $\rho_c$ & \textit{T$_{max}$} & \textit{T$_c$-T$_{max}$} & \textit{T$_g$} \\
& (K) & (g cm$^{-3}$) & (K) & (K) & (K)\\
\noalign{\smallskip}\hline\noalign{\smallskip}
ST2 & 550.2 & 0.286 & 305 & 246 &\ 235\\
 & & & $\sim$300 \cite{ST2} & & \\
ST2RF& 536.2 & 0.286	& 320 & 216 &\	255\\
 &&& $\sim$330 \cite{Har97} & &\\
TIP4P&	580 & 0.328	& $\sim$ 250	& 330 &\	180 \\
&579 \cite{Zeng02}&&263 \cite{Saitta}&&\\
&588.2 \cite{Nez04}& 0.315 \cite{Nez04} &&&\\
TIP5P&537.7&0.290&$\sim$ 275 & 263 &\ 215 \\
&521.3 \cite{Nez04}&0.337\cite{Nez04} &\ 275 \cite{TIP5P}&&\\
&546 \cite{Zeng02}&&&&\\
SPCE&623.3&0.319&&& 220\\
&640\cite{GG93} & 0.29 \cite{GG93}&235 \cite{Baez94}&&\ 177 \cite{Baez94}\\
&630\cite{Econ98} & 0.295 \cite{Econ98}&255 \cite{Baez95}&&$\sim$210 \cite{Baez95}\\
&638.6\cite{Pan98}& 0.273 \cite{Pan98}&$\sim$240 \cite{Har97}&&\  188 \cite{Gi04}\\
Exp. &647 \cite{CCexp}&0.326 \cite{CCexp}&277.1 \cite{CCexp}&374.6&$\sim$160 \cite
{Velikov}\\
&&&&&$\sim$136 \cite{Angellrev}\\
\noalign{\smallskip}\hline
\end{tabular}
\end{table}   
\section{\label{sec:Results}Results}
\subsection{\label{sec:CClv}Liquid-vapor coexistence of water from the critical point to the glassy state}

\subsubsection{\label{sec:CCro}Liquid density along the liquid-vapor coexistence curve} 
The liquid-vapor coexistence curves of the ST2 and ST2RF water models are shown in Fig.1. At high temperature in both models the liquid densities are essentially lower than the experimental values due to the strongly underestimated critical temperatures \textit{T$_c$} (see Table I). At ambient temperatures the liquid density of the ST2 model is close to the density of real water, whereas in the ST2RF model it remains significantly underestimated (Fig.2). Our results of the GEMC simulations (Fig.2) agree well with the results of molecular dynamics (MD) simulations of the ST2 \cite{ST2} and ST2RF \cite{Poole93a} models at \textit{P} $\approx$ 0. The ST2 model shows the density maximum at about 305 K and well reproduces the experimental liquid density in a temperature range of about 70$^{\circ}$, when the latter is shifted upwards by 28$^{\circ}$ (see Fig.2). At all temperatures the liquid density of the ST2RF model is significantly below the experimental values. In the ST2RF model the temperature \textit{T$_{max}$} of the density maximum (about 320 K) is slightly higher than in the ST2 water, and  the absolute value of the liquid density $\rho_l$ is about 7$\%$ below the experimental value. The experimental density $\rho_l$  of about 1 g cm$^{-3}$ can be reproduced in the ST2RF model by applying a positive pressure of about 800 bars.\cite{Poole92,Paschek} The drastic density difference between ST2 and ST2RF models is caused by the application of the reaction field method for the treatment of the LRCI corrections in the latter case, which apparently strengthens the water hydrogen bonding. The shift of the liquid density in the ST2RF model agrees with the predictions of the modified van der Waals model for water.\cite{Hbonds}  
The characteristic temperature interval from \textit{T$_c$} to \textit{T$_{max}$} is only 66$\%$ of experimental value for the ST2 model and only 58$\%$ for the ST2RF model (Table I). Note, that with decreasing temperature a density minimum follows the density maximum in both models (see Figs.1 and 2). 
\begin{figure}[t]
\includegraphics[width=6cm]{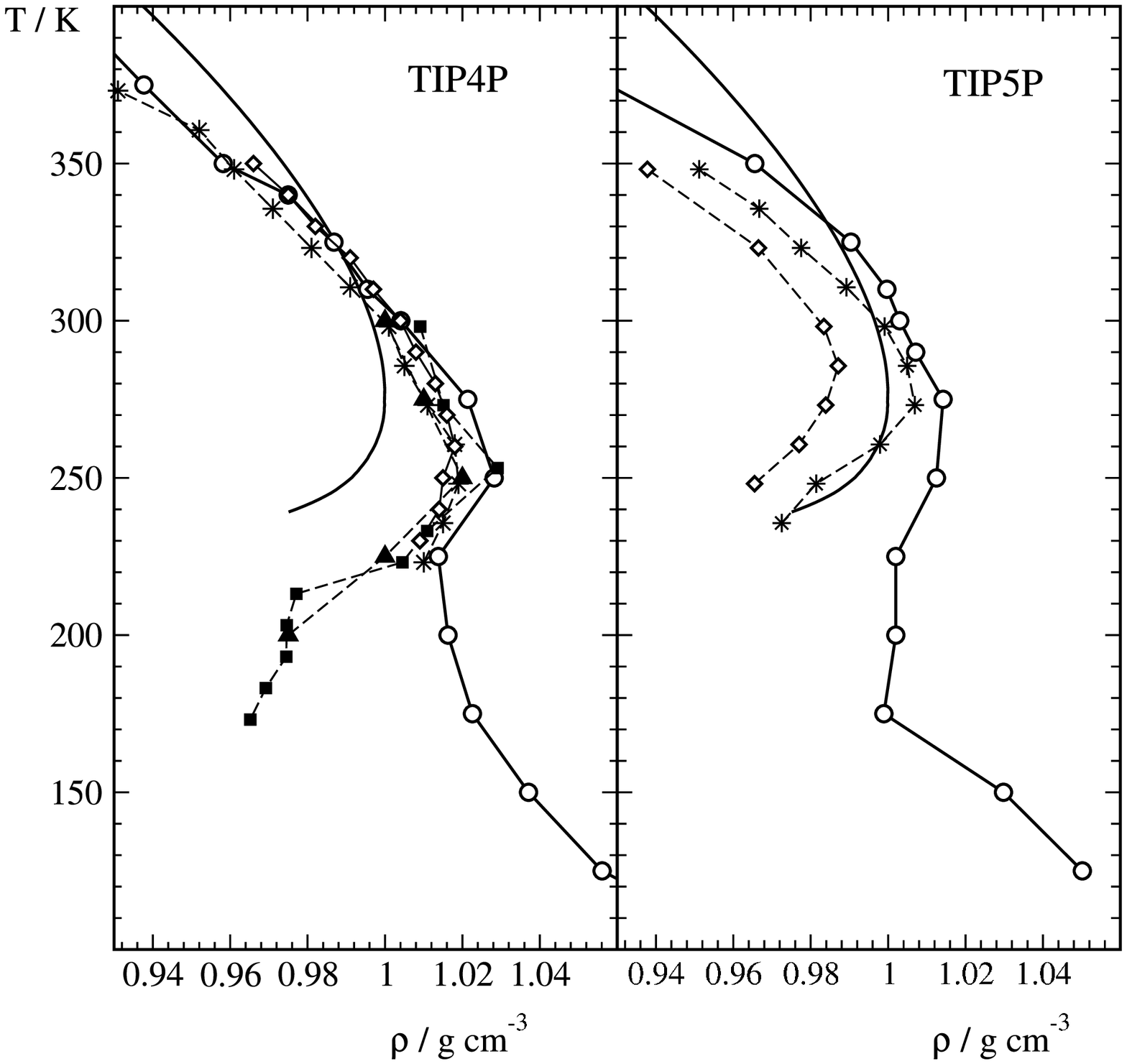}
\caption{ Liquid densities of TIP4P and TIP5P water models along the liquid-vapor coexistence curve (open circles). Simulation data at \textit{P} $\approx$ 0  for the TIP4P model: squares, \cite{Tanaka00} triangles, \cite{Poole93a} stars, \cite{Jorg98} diamonds. \cite{Saitta} Simulation data at P $\approx$ 0 for the TIP5P model: stars,\cite{TIP5P} diamonds.\cite{Nez02} Experimental data \cite{CCexp,CCexpZ} are shown by solid lines.}
\end{figure} 
\par
The striking feature of the liquid-vapor coexistence curve of the ST2 model is a break of the liquid branch at \textit{T} = 270 K (Figs.1 and 2). At this temperature two liquid phases with different densities (about 0.91 and 0.97 g cm$^{-3}$) coexist with the vapor phase. This is not the case for any other studied temperatures, including the nearest ones \textit{T} = 265 K and \textit{T} = 275 K, where only one liquid phase coexists with the vapor. Evidently, there is a triple point at \textit{T} = 270 K, where two first order phase transitions (liquid-vapor and liquid-liquid) meet each other. The gradual change of $\rho_l$ in the whole studies temperature range (Fig.2) indicates the absence of the triple point in the ST2RF model. The disappearance of the triple point with the strengthening of the water hydrogen bonding qualitatively agrees with the predictions of the modified van der Waals model for water. \cite{Hbonds}  
\begin{figure}[t]
\includegraphics[width=6cm]{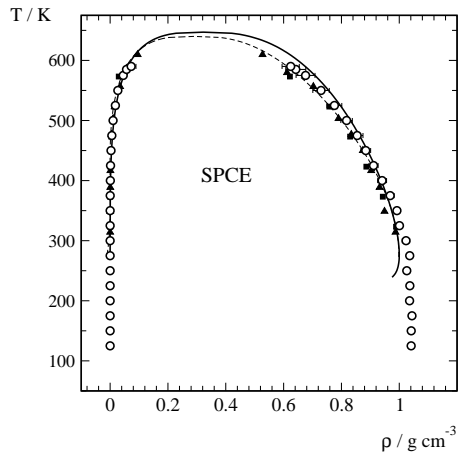}
\caption{ Liquid-vapor coexistence curve of SPCE water model (open circles). Simulation data from Ref.\onlinecite{GG93}, Ref.\onlinecite{Econ98} and Ref.\onlinecite{Maurer00} are shown by dashed line, solid triangles and solid squares, respectively. Experimental data \cite{CCexp,CCexpZ} are shown by solid line.}
\end{figure} 
\par
The liquid-vapor coexistence curves of the TIP4P and TIP5P water models are compared with the experimental data in Fig.3. Upon cooling the liquid density in both models passes through a maximum and afterward through a minimum. Both models underestimate the critical temperature (Table I), that essentially depresses the liquid density at high temperatures, especially in the case of the TIP5P model. For the TIP4P water model the characteristic temperature interval from \textit{T$_c$} to \textit{T$_{max}$} is about 88$\%$ of the experimental value, whereas for the TIP5P model it is only 70$\%$, which is close to the value for another five-site water model, the ST2 model (see Table I). Most of the data points for the TIP4P model are from our previous paper, \cite{BGO2004} where comparisons with other simulation studies of the coexistence curve can be found (see Fig.5 in Ref.\onlinecite{BGO2004}). Depending on the implementation of the TIP4P  and TIP5P water models (long-range corrections for intermolecular interactions, cutoff, system size, etc.), the density of the saturated liquid at room temperature varies within 1-2 $\%$. In particular, using the LRCI corrections causes a slight decrease of the liquid density, whereas including the LRLJ corrections causes the opposite trend. 
\par
The details of the particular implementation of the water model and the employed simulation method gain more importance with decreasing temperature (Fig.4). At ambient temperatures the liquid density of the TIP4P water model is not very sensitive to the simulation details. However, two qualitatively different courses of the liquid density are obtained in the supercooled region: in our simulation $\rho_l$ starts to increase with decreasing temperature at \textit{T} $<$ 225 K, whereas in other simulations \cite{Poole93a,Tanaka00} it continues to decrease. Apparently, this difference originate from the use of LRLJ corrections in the former case and LRCI corrections in the latter case. 

\par
The TIP5P water model seems to be even more sensitive to the simulation details. \cite{Nez02,Rick04,Pitera04} In the region of the density maximum the use of LRCI corrections  \cite{Nez02} causes a decrease of $\rho_l$ by about 2$\%$, whereas the use of LRLJ corrections (our simulations) causes a slight increase of $\rho_l$ in comparison with the original TIP5P model,\cite{TIP5P} which was parameterized without any long-range corrections for intermolecular interactions (Fig.4). In the supercooled region the increase of $\rho_l$ due to the LRLJ corrections achieves about 3$\%$. 
\begin{figure}[t]
\includegraphics[width=5cm]{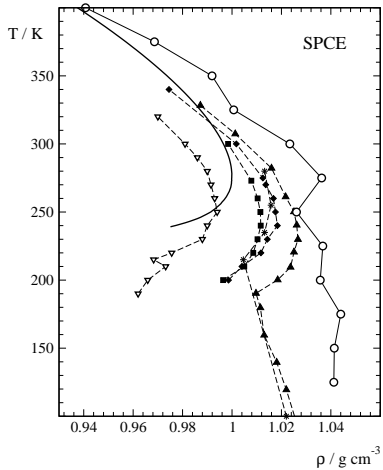}
\caption{ Liquid densities of the SPCE water models along the liquid-vapor coexistence curve (open circles). Simulation data at \textit{P} $\approx$ 0: triangles up,\cite{Baez94} triangles down, \cite{Baez95} squares, \cite{Har97} stars \cite{Bel99} and diamonds. \cite{Bryk} Experimental data \cite{CCexp,CCexpZ} are shown by the solid line.}
\end{figure} 
\par
Among the considered water models, the SPCE model gives the most realistic value of \textit{T$_c$} (see Table I), and accordingly the closeness of the simulated and experimental values of the coexisting densities at high temperatures (Fig.5). The characteristic temperature interval from T$_c$ to T$_{max}$ is also close to the experimental value (Table I). Application of the LRCI corrections \cite{GG93,Econ98,Maurer00} depresses the liquid density by about 2-3$\%$ similarly to other water models. At low temperatures the behavior of the density of the saturated liquid in the SPCE model seems to be more sensitive to the simulation details in comparison with other models (Fig.6). In the region of the density maximum the value of $\rho_l$ is the largest, when only LRLJ corrections are taken into account (Fig.6, circles). It subsequently decreases, when such corrections are excluded (Fig.6, solid triangles), when both kinds of the long-range corrections are included (Fig.6, diamonds) and when only LRCI corrections are included (Fig.6, squares). The use of a smaller spherical cutoff results in the lowest liquid density (Fig.6, open triangles).
\begin{figure}[t]
\includegraphics[width=6cm]{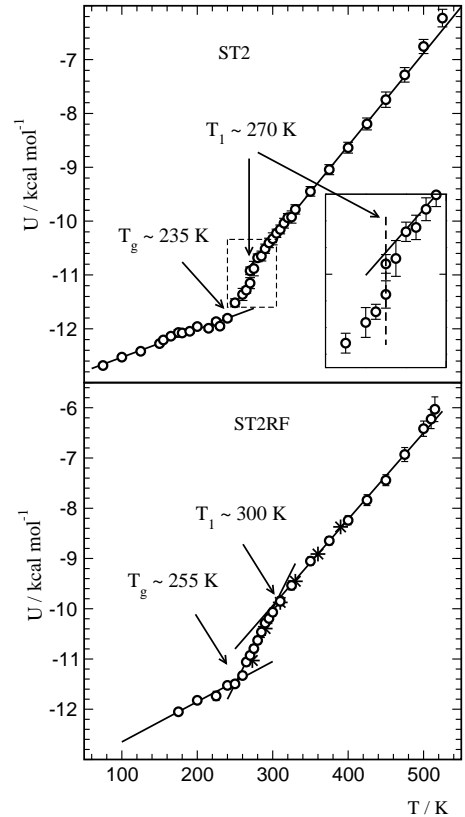}
\caption{Potential energy \textit{U} in the liquid phase along the liquid-vapor coexistence curve of the ST2 and ST2RF water models (open circles). Data points in the vicinity of the liquid-liquid transition of the ST2 model at \textit{T} $\approx$ 270 K are shown in the inset. The values of \textit{U}, obtained from simulations of the ST2RF model at \textit{P} $\approx$ 0,\cite{Poole93a} are shown by stars.}
\end{figure} 
\par
Some features of the temperature dependence of the saturated liquid density at ambient and supercooled temperatures seem to be general for all studied water models, despite the high sensitivity of the phase diagram to the model and its implementation. All studied water models show the well-known density maximum (see Table I). This maximum is followed by a density minimum with decreasing temperature. Previously, we observed the liquid density minimum along the liquid-vapor coexistence curve for TIP4P water \cite{BGOPCCP,NATO,BGO2004} (Figs.3 and 4), ST2 water \cite{BGO03} (Figs.1 and 2) and TIP5P water \cite{Kyoto} (Figs.3 and 4). Now, the ST2RF water model also shows a pronounced density minimum (Figs. 1 and 2). This well agrees with the recent observation of a density minimum of this model at negative pressures. \cite{Poole05} Although we are not able to locate both a density minimum and a maximum in our simulations of SPCE water (Fig.5), the existence of the density minimum seems to be quite possible for other implementations of this model (Fig.6). Note, that the liquid density minimum around 185-195 K was also observed upon temperature quench of a water model, proposed by Guillot and Guissani. \cite{GG2003}    
\subsubsection{\label{sec:glass} Potential energy of liquid water along the liquid-vapor coexistence curve and glass transition} 
The temperature dependence of the potential energy \textit{U} of the water molecules talong the liquid-vapor coexistence curve is shown in Fig.7 for the ST2 model. The discontinuity of \textit{U} (of about 0.23 kcal/mol) at \textit{T} = 270 K (see inset in Fig.7) reflects the liquid-liquid phase transition at the triple point. The slope of \textit{U}(\textit{T}) does not change noticeably when crossing the triple point, indicating the essentially liquid nature of the coexisting phases. A slower change of \textit{U}(\textit{T}), typical for the glassy state, is observed at low temperatures. This allows an estimate of the glass transition temperature of the ST2 water model at zero pressure as \textit{T$_g$} $\approx$ 235 K.
\begin{figure}[t]
\includegraphics[width=6cm]{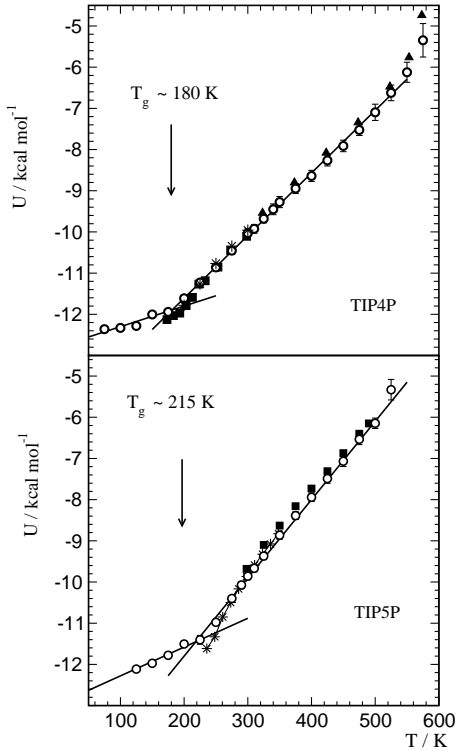}
\caption{ Potential energy \textit{U} in the liquid phase along the liquid-vapor coexistence curve of the TIP4P and TIP5P water models (open circles). The values of \textit{U}, obtained from simulations at \textit{P} $\approx$ 0 are shown by squares,\cite{Tanaka00} stars \cite{Poole93a} and triangles.\cite{Maurer00}}
\end{figure} 
\par
The temperature dependence of the potential energy \textit{U} of the STRF water model indicates a transition to the glassy state at \textit{T$_g$} $\approx$ 255 K, i.e. slightly higher then in the ST2 model. Two different slopes of \textit{U}\textit{(T)} above \textit{T$_g$ }evidences two quite different kinds of liquid water. The continuous transition between these two liquids occurs at about 300 K, i.e. about 20$^{\circ}$ below the temperature of the density maximum (Fig.7, Table I).  

\par
Transitions to glassy states are also seen for the TIP4P and TIP5P water models (Fig.8). The estimated values of the glass transition temperature are: \textit{T$_g$} $\approx$ 180 K for the TIP4P model and \textit{T$_g$} $\approx$ 215 K for the TIP5P model. Note, that the use of the LRCI correction (Fig.8, squares and stars in the upper panel and stars in the lower panel) results in noticeably lower potential energies at \textit{T} $\leq$ 225 K for the TIP4P model and \textit{T} $\leq$ 275 K for the TIP5P model. This is obviously connected with the essentially different densities of water in the two implementations of the TIP4P and TIP5P water models in these temperature ranges (see Fig.4). The glass transition temperature of the SPCE water model noticeably depends on its implementation (see Table I) and from our simulation study we estimate glass transition temperature \textit{T$_g$} $\approx$ 220 K for this model (Fig.9).
\subsubsection{\label{sec:CCucp} Heat capacity along the liquid-vapor coexistence curve}
The heat capacity at constant pressure \textit{C$_p$ }is the temperature derivative of the enthalpy and relates to the potential energy of a system through the equation:
\begin{eqnarray}
\textit{C}_\textit{p} = \left(\frac{\partial\left(\textit{U} + \textit{PV} + 3\textit{RT}\right)}{\partial \textit{T}}\right)_\textit{P}. 
\end{eqnarray}   
Along the liquid-vapor coexistence curve, the contribution of \textit{P}($\partial$ V/$\partial$ T)$_P$ to \textit{C$_p$} is negligible up to temperatures of about 450 to 500 K. We have computed \textit{C$_p$} from 2- or 3-point running averages of the enthalpy \textit{\textit{U} + 3\textit{RT}} followed by central differentiation. The data for the lowest temperature was estimated using the forward formula for the derivative.
\begin{figure}[t]
\includegraphics[width=6cm]{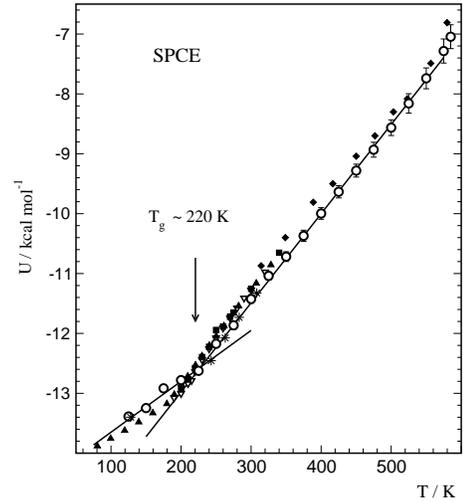}
\caption{Potential energy \textit{U} in the liquid phase along the liquid-vapor coexistence curve of the SPCE water model (open circles). Values of \textit{U}, obtained from simulations at \textit{P} $\approx$ 0: triangles up,\cite{Baez94} triangles down,\cite{Baez95} squares, \cite{Har97} stars \cite{Bel99} and diamonds. \cite{Econ98}}
\end{figure} 
\par
Our simulation results are compared with the experimental data and other simulation studies of \textit{C$_p$} in Figs.10 and 11. The accuracy of the obtained \textit{C$_p$} data directly depends on the number of simulated temperature points. The available simulation data allow an estimation of the average value of \textit{C$_p$} over a wide temperature interval, whereas the temperature dependence of \textit{C$_p$} can be analyzed only qualitatively. In some temperature intervals the temperature dependence of \textit{U} is also approximated by a linear behavior (see solid lines in Figs.7 to 9) and the corresponding average values of \textit{C$_p$} are shown by horizontal dashed lines in Figs.10 and 11. The value of \textit{C$_p$} in the high-temperature liquid is about twice the \textit{C$_p$} in the glassy state. 
\begin{figure}[t]
\includegraphics[width=5cm]{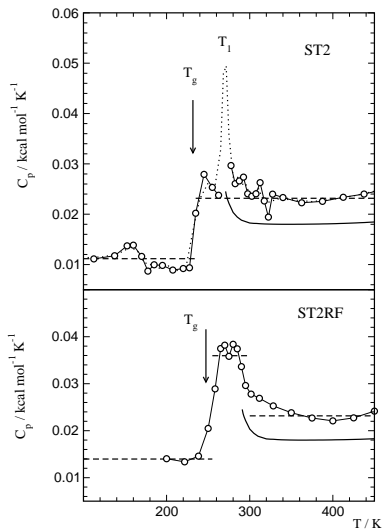}
\caption{ Heat capacity of liquid water along the liquid-vapor coexistence curve: our simulation (open circles) and experiment \cite{heat1,CCexp} (thick lines). Experimental data are shifted by +35$^{\circ}$ and +55$^{\circ}$ for comparison with the ST2 and ST2RF water models, respectively. Dotted line shows an estimate of \textit{C$_p$}, when the discontinuity at the liquid-liquid transition is not taken into account. Dashed lines indicate the values of \textit{C$_p$}, estimated from the linear approximations of \textit{U}, shown by solid lines in Fig.7. }
\end{figure}
\par
Different temperature dependences of \textit{C$_p$} are observed for the different water models. An increase of the heat capacity of the ST2 water model is clearly seen with decreasing temperature down to 270 K, i.e. to the triple point of the liquid-vapor and liquid-liquid phase transition (Fig.10, upper panel). An estimation of \textit{C$_p$}, which neglects the discontinuous drop of the potential energy \textit{U} at \textit{T}$_1$ = 270 K (see inset of Fig.7), results in a much stronger apparent increase of the heat capacity (dotted line in Fig.10). Such an artificially strong increase of the estimated value of \textit{C$_p$} could appear, if the liquid-liquid phase transition is smeared our due to the limitations of the simulation method. The heat capacity of the low temperature liquid phase can hardly be estimated due to the small temperature interval between the liquid-liquid phase transition and the glass transition (see Fig.7), but seems to be of the same magnitude as above the triple point. 

\par
The ST2RF model also shows an increase of \textit{C$_p$} upon cooling (Fig.10, lower panel). At about 300 K, where a gradual transition to the low density form of liquid water occurs, \textit{C$_p$} increases abruptly to a value of about 0.04 kcal mol$^{-1}$ K$^{-1}$, and remains at such high values till the transition to the glassy state. So, the heat capacity of the second (low-density, low-temperature) liquid in the ST2RF model is almost two times higher, than the heat capacity of the ordinary liquid, which coexists with the vapor in a wide temperature range up to the liquid-vapor critical point.
\par
Similarly to the ST2 and ST2RF models, the TIP5P water model shows a noticeable increase of \textit{C$_p$} with decreasing temperature (Fig.11, middle panel). This behavior is interrupted by the transition to a glassy state. The absolute values of \textit{C$_p$} in the TIP4P and SPCE water models are rather close to the experimental data in a wide temperature range (Fig.11). However, these models do not show the noticeable increase of \textit{C$_p$} upon cooling, observed in real water. 
\begin{figure}[t]
\includegraphics[width=5cm]{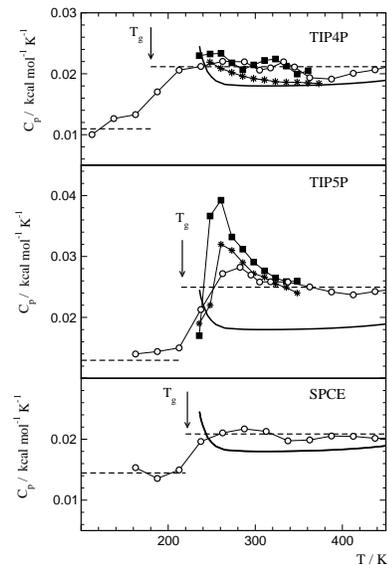}
\caption{ Heat capacity of liquid water along the liquid-vapor coexistence curve: our simulation (open circles) and experiment \cite{heat1,CCexp} (thick lines). Data from other simulation studies at \textit{P} $\approx$ 0: TIP4P model \cite{Jorg98} (squares, upper panel), TIP4P-Ew model \cite{Pitera04} (stars, upper panel), TIP5P model \cite{TIP5P} (squares, middle panel), TIP5P-Ew model \cite{Rick04} (stars, middle panel). Dashed lines indicate the values of \textit{C$_p$}, estimated from the linear approximations of \textit{U}, shown by solid lines in Figs.8 and 9.}
\end{figure}
\subsection{\label{sec:CCll}Liquid-liquid phase transitions of water at low temperatures} 
\subsubsection{\label{sec:ST2}ST2 model} 
Isotherms of the ST2 water model, obtained by MC simulations in the restricted \textit{NPT} ensemble at \textit{T} = 235 to 290 K, are shown in Fig.12. No liquid-liquid transitions could be detected at \textit{T} = 290 K, where the liquid density gradually changes with pressure. At \textit{T} = 275 K a liquid-liquid coexistence appears at slightly negative pressure (\textit{P} $\approx$ -0.2 kbar), the densities of the coexisting liquid phases are about $\sim$0.94 and $\sim$0.98 g cm$^{-3}$. At \textit{T} = 260 K this transition shifts to positive pressure (\textit{P} $\approx$ +0.4 kbar) and the two-phase region becomes wider (from about 0.90 to 1.01 g cm$^{-3}$). So, at some temperature between 275 K and 260 K this liquid-liquid transition should pass through zero pressure, i.e. should meet the liquid-vapor phase transition at a triple point. Indeed, such a triple point at \textit{T} = 270 K was found at the liquid-vapor coexistence curve of the ST2 model (Figs.1 and 2). The densities of the coexisting liquid phases at the triple point, estimated from the GEMC simulations of the liquid-vapor coexistence, $\sim$0.91 and $\sim$0.97 g cm$^{-3}$, are in good agreement with the MC simulations in the restricted \textit{NPT}ensemble. Moreover, extensive direct equilibration of the two liquid phases in the Gibbs ensemble also gives a rather similar coexistence region: between 0.92 and 1.00 g cm$^{-3}$ at \textit{T} = 270 K and between 0.90 and 1.02 g cm$^{-3}$ at \textit{T} = 260 K (see Fig.4 in Ref.\onlinecite{BGO03}). Pressure and densities of these coexisting phases do not change noticeably upon cooling to \textit{T} = 235 K. 
\begin{figure}[t]
\includegraphics[width=6cm]{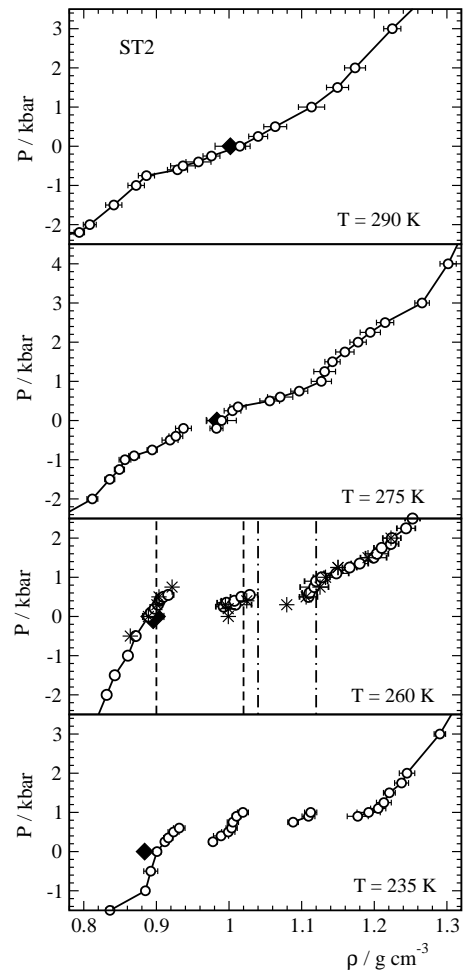}
\caption{ Isotherms of ST2 water from simulations in the restricted ensemble with 513 (open circles) and 621 (stars) water molecules. The density of the liquid phase at the liquid-vapor coexistence curve (\textit{P} = 0 bar) is shown by diamonds. The densities of the coexisting phases of the first and second liquid-liquid phase transitions of ST2 water at 260 K, obtained by GEMC simulations, are shown by dashed and dot-dashed lines, respectively. }
\end{figure} 
\par
At \textit{T} = 260 K the ST2 water model shows a second liquid-liquid phase transition, occurring at positive pressure (\textit{P} $\approx$ +0.5 kbar) in the density interval from $\sim$1.02 to $\sim$1.11 g cm$^{-3}$. This interval agrees well with the densities of the coexisting phases, obtained directly in GEMC simulations: 1.03 and 1.12 g cm$^{-3}$ at \textit{T} = 270 K and 1.04 and 1.12 g cm$^{-3}$ at \textit{T} = 260 K (Fig.4 in Ref.\onlinecite{BGO03}). The densities of the coexisting phases do not change noticeably at \textit{T} = 235 K, whereas the coexistence pressure slightly increases up to \textit{P} $\approx$ +0.85 kbar. A third liquid-liquid phase transition of the ST2 water model appears at \textit{T} = 235 K. It is located at positive pressure (\textit{P} $\approx$ +0.95 kbar) (Fig.12) and the densities of the coexisting liquid phases are $\sim$1.12 and $\sim$1.19 g cm$^{-3}$. 

\par
The location of the thus observed liquid-liquid coexistence regions with respect to the liquid branch of the liquid-vapor coexistence curve is shown schematically in upper panel of Fig.13 (see also Fig.4 of Ref.\onlinecite{BGO03}. The simulation results evidence, that the first (the lowest-density) liquid-liquid phase transition of the ST2 model should have a critical point of demixing at slightly negative pressures with a critical temperature between 275 K and 290 K and a critical density between $\sim$0.94 and $\sim$0.98 g cm$^{-3}$. It seems likely, that the second and third phase transitions also end at corresponding critical points. However, the observed temperature evolution of the coexisting densities of these transitions does not allow to exclude the possibility of a common critical point. If so, the narrow one-phase region near 1.1 g cm$^{-3}$ should end in a triple point, where three liquid phases coexist. Note, that all three phase transitions occur above the glass transition temperature \textit{T$_g$} $\approx$ 235 K, estimated at zero pressure. This means, that the first transition is definitely a transition between two \textit{liquid} phases. If \textit{T$_g$} does not strongly increase with pressure, the two other transitions also should be essentially liquid-liquid. 

 \subsubsection{\label{sec:ST2RR}ST2RF model} 
 Several isotherms of the ST2RF water model, obtained by MC simulations in the restricted NPT ensemble at \textit{T} = 200 to 290 K, are shown in Figs.14 and 15 together with the results of other simulation studies. There is good agreement of our data for the supercritical isotherms with MD simulations in the \textit{NVT} ensemble (Fig.14, \textit{T} = 290 and 275 K). In this water model the first liquid-liquid phase transition appears at \textit{T} = 260 K and \textit{P} $\approx$ +1.3 kbar in the density interval from $\sim$0.94 to $\sim$1.01 g cm$^{-3}$. The shape of the isotherm indicates a critical point at positive pressures slightly above 260 K. 
The shift of the critical point from negative pressure for the ST2 model to positive pressure for the ST2RF model and the subsequent disappearance of the triple point corroborates with the changes of the water phase diagram expected due to strengthening of the hydrogen bonding.\cite{Hbonds}
 At \textit{T} = 235 K the two-phase region becomes wider (from $\sim$0.94 to $\sim$1.10 g cm$^{-3}$) and the coexistence pressure increases to \textit{P} $\approx$ +2.0 kbar. Our isotherm at \textit{T} = 235 K agrees well with the results of other simulation studies (Fig.15). The systematic shift of the data points, obtained from a sedimentation profile, \cite{Sciort03} toward higher pressures is caused by a slightly lower temperature (230 K) and the absence of the LRLJ corrections in that study.  
\par
 The first liquid-liquid phase transition of the ST2RF model is also clearly seen in the \textit{T} = 200 K isotherm, i.e. deeply below the glass transition temperature \textit{T$_g$} $\approx$ 255 K, estimated at zero pressure. Probably due to the fact that the 200 K isotherm refers to a glassy state the different branches of the isotherm start to overlap in some density range (Fig.14, lower panel). This substantially complicates estimating the coexisting densities and the pressure at the transition. However, this transition remains roughly in the same pressure and density interval, as at \textit{T} = 235 K. 

\par
At \textit{T} = 200 K the ST2RF model shows a pronounced second liquid-liquid (or better amorphous-amorphous) transition at \textit{P} $\approx$ +3 kbar with a two-phase region roughly from 1.1 to 1.2 g cm$^{-3}$. The location of the observed coexistence regions with respect to the liquid branch of the liquid-vapor coexistence curve of the ST2RF model is shown schematically in Fig.13 (lower panel). Note, that the density intervals of the second transition of the ST2RF model and the third transition of the ST2 model are very close. The coexistence region of the first transition of the ST2RF model covers the two coexistence regions of the first and second transitions of the ST2 model.  
\begin{figure}[t]
\includegraphics[width=6cm]{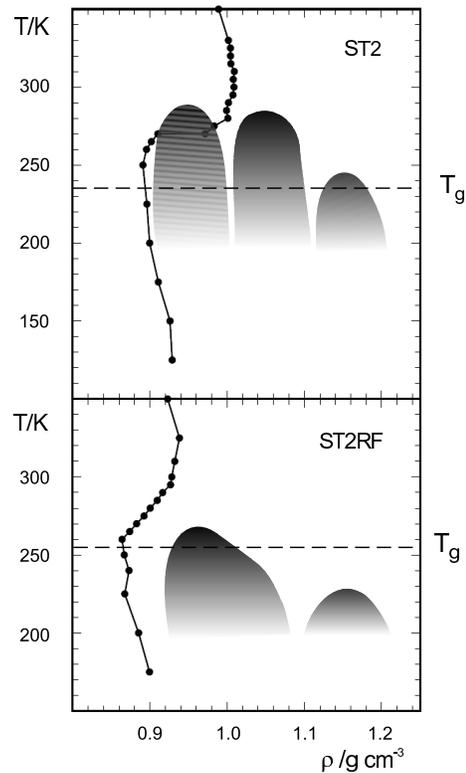}
\caption{Schematic representation of the location of the liquid-liquid phase transitions (shadowed areas) with respect to the liquid branch of the liquid-vapor coexistence curve (solid circles) for the ST2 and ST2RF water models. Glass temperature at \textit{P} = 0 is indicated by dashed line.}
\end{figure} 
\subsubsection{\label{sec:TIP4P}TIP4P model} 
The \textit{T} = 200 K isotherm of the TIP4P water model shows a liquid-liquid phase transition  in the density interval from $\sim$1.05 to $\sim$1.08 g cm$^{-3}$ (Fig.16). The critical point of this transition should be located slightly above 200 K, at \textit{P} $\approx$ +1 kbar and $\rho$ $\approx$ 1.06 g cm$^{-3}$. Continuous isotherms of TIP4P water were obtained in MD simulations using the \textit{NVT} \cite{Poole93a} and \textit{NPT} \cite{Tanaka96} ensembles (see stars and triangles, respectively, in Fig.16). The shift of the MD data to lower densities may be attributed to the use of LRCI corrections in Refs.\onlinecite{Poole93a} and \onlinecite{Tanaka96}. The compressibility maximum, obtained in the MD simulations \cite{Poole93a,Poole97a} at $\rho$ $\approx$ 1.03 g cm$^{-3}$ is rather close to the critical density at 1.06 g cm$^{-3}$ in our simulations. The absence of the liquid-liquid transition at \textit{T} = 200 K in the MD simulation studies \cite{Poole93a,Tanaka96} should be attributed to the limitations of the applied simulation methods, which underestimate the critical temperature (see Introduction).   
\begin{figure}[t]
\includegraphics[width=6cm]{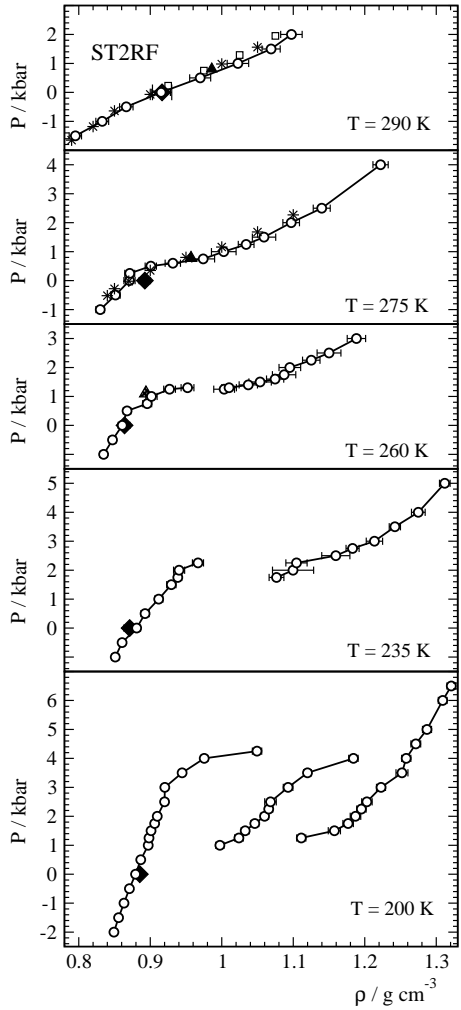}
\caption{ Isotherms of ST2RF water from MC simulations in the restricted \textit{NPT} ensemble (open circles) and densities of the liquid phase at the liquid-vapor coexistence curve (diamonds at \textit{P} = 0 bar). Data points from MD simulations in the \textit{NVT} ensemble are shown by stars, \cite{Poole93a} open squares \cite{Har97a} and solid triangles. \cite{Paschek}}
\end{figure}
\par
When decreasing the temperature to 175 K, two liquid-liquid phase transitions are seen. The first one is located at \textit{P} $\approx$ -0.8 kbar with densities of the coexisting phases of 0.96 and 1.02 g cm$^{-3}$. This transition is located in the region, which is metastable with respect to evaporation (see diamonds in Fig.16). Further decrease of the temperature to \textit{T} = 150 K, which is essentially below the glass transition temperature \textit{T$_g$} $\approx$ 180 K, estimated at zero pressure, the different branches of the isotherm overlap in a wide range of pressure and density (Fig.16, lower panel). At \textit{T} = 150 K the first transition apparently remains at negative pressure and its parameters (center of the overlap region) seem to be close to those at \textit{T} = 175 K. At \textit{T} = 175 K the second liquid-liquid transition of the TIP4P water model occurs at P $\approx$ +1.8 kbar with densities of the coexisting phases of 1.08 and 1.14 g cm$^{-3}$ (Fig.16, middle panel). When decreasing the temperature to 150 K, this two-phase region becomes wider and extends up to 1.2 g cm$^{-3}$ (Fig.16, lower panel).   
\par
The liquid branch of the liquid-vapor coexistence curve of the TIP4P model varies smoothly with temperature and does not show any triple point (Figs.3, 4). Therefore, the first liquid-liquid transition, which occurs at negative pressure at \textit{T} = 150 and 175 K, should end at a critical point between 175 and 200 K also at negative pressure. 
 Isotherms of TIP5P water from MC simulations in the restricted NPT ensemble (open circles) and densities of the liquid phase at the liquid-vapor coexistence curve (diamonds at \textit{P} = 0 bar). Data points from MD simulations in the \textit{NVT} ensemble \cite{Sciort02} are shown by stars (for \textit{T} = 225 K these points were obtained by interpolation of the data for \textit{T} = 220 and 230 K).

\par
 The location of the two liquid-liquid transitions with respect to the liquid branch of the liquid-vapor coexistence curve of the TIP4P model is shown schematically in Fig.17. The results of MD simulations of supercooled TIP4P water \cite{Poole93a,Tanaka96,Poole97a} support this picture. Indeed, the drop of the water density from about 1.12 to 1.05 g cm$^{-3}$ between 193 and 173 K along the isobar \textit{P} = 2 kbar in Ref.\onlinecite{Tanaka96}, can be understood as the crossing of the second liquid-liquid transition of TIP4P water (Fig.17). The sharp change of the water density along the isobar \textit{P} $\approx$ 0 \cite{Tanaka96} probably reflects the proximity of the first liquid-liquid phase transition, which may be located at slightly positive pressures in that implementation of the TIP4P model (see squares in Fig.4).   
\subsubsection{\label{sec:TIP5P}TIP5P model}
Isotherms of the TIP5P water model at supercooled temperatures are shown in Fig.18. The inflection point of the \textit{T} = 250 K isotherm of the TIP5P model is located approximately at the same pressure and density, as the inflection point of the \textit{T} = 200 K isotherm of the TIP4P model (Fig.16). Note, that a similar shape of this isotherm was obtained from MD simulations in the \textit{NVT} ensemble.\cite{Sciort02} At \textit{T} = 225 K a liquid-liquid phase transition appears at \textit{P} $\approx$ +1.3 kbar with densities of the coexisting phases $\sim$1.03 and $\sim$1.08 g cm$^{-3}$. This transition was not observed at this temperature in MD simulations,\cite{Sciort02} probably due to the unavoidable distortion of the subcritical isotherms in the \textit{NVT }ensemble. With decreasing temperature the coexistence pressure of this transition increases noticeably and the densities of both coexisting phases shift to higher values. 
\par
A lower-density liquid-liquid transition appears at \textit{T} = 175 K with densities of the coexisting phases $\sim$0.99 and $\sim$1.03 g cm$^{-3}$. Obviously, this transition appears at slightly positive pressure (see the location of the liquid density at the liquid-vapor coexistence curve at \textit{T} = 175 K indicated by a diamond in Fig.18). At \textit{T} = 150 K this transition is observed at negative pressures (about -3 kbar) in the density interval from $\sim$0.94 to $\sim$0.99 g cm$^{-3}$. Thus, this liquid-liquid transition should cross the liquid-vapor coexistence curve at a triple point between 150 and 175 K. Indeed, a sharp change of the liquid density at the liquid-vapor coexistence curve is observed at this temperature interval (see Fig.4). The location of the estimated liquid-liquid two-phase regions with the respect to the liquid branch of the liquid-vapor coexistence curve of the TIP5P model is shown schematically in Fig.17 (middle panel). Note some discrepancy between this phase diagram and the liquid-liquid phase transition with a critical point at \textit{T} = 217 K, \textit{P} = 3.4 kbar and $\rho$ = 1.13 g cm$^{-3}$, estimated for the TIP5P model in Ref. \onlinecite{Sciort02}.
\subsubsection{\label{sec:SPCE}SPCE model} 
The appearance of a liquid-liquid (or better amorphous-amorphous) phase transition of the SPCE model is seen from the isotherm \textit{T} = 200 K (Fig.19). The critical point of this transition we estimate at temperature slightly above 200 K, \textit{P} $\approx$ +0.9 kbar and $\rho$ $\approx$ 1.06 g cm$^{-3}$. With decreasing temperature the coexistence pressure of this transition remains positive, whereas the density range of the two-phase coexistence increases from 1.07 - 1.13 g cm$^{-3}$ at \textit{T} = 175 K to 1.09 - 1.16 g cm$^{-3}$ at \textit{T} = 150 K. A lower-density amorphous-amorphous phase transition appears at slightly negative pressure, the coexistence region we estimate as 0.99 - 1.04 g cm$^{-3}$ at \textit{T} = 175 K and 0.97 - 1.03 g cm$^{-3}$ at \textit{T} = 150 K. The location of these two liquid-liquid transitions with respect to the liquid branch of the liquid-vapor coexistence curve is shown schematically in Fig.17. Studies of the liquid-liquid transition of the SPCE model, based on an analytic expansion for the liquid free energy \cite{Scala} do not agree with our phase diagram. First, the existence of a \textit{single} liquid-liquid transition was imposed in Ref.\onlinecite{Scala}. The critical point of this single liquid-liquid transition, estimated at \textit{T} $\approx$ 130 K \textit{P} $\approx$ +2.9 kbar and $\rho$ $\approx$ 1.10 g cm$^{-3}$, is essentially shifted to lower temperature and higher pressure in comparison with the critical points of the two liquid-liquid phase transitions, shown in Fig.17. This shift is probably caused by the use of LRCI corrections in Ref.\onlinecite{Scala}. 
\begin{figure}[t]
\includegraphics[width=6cm]{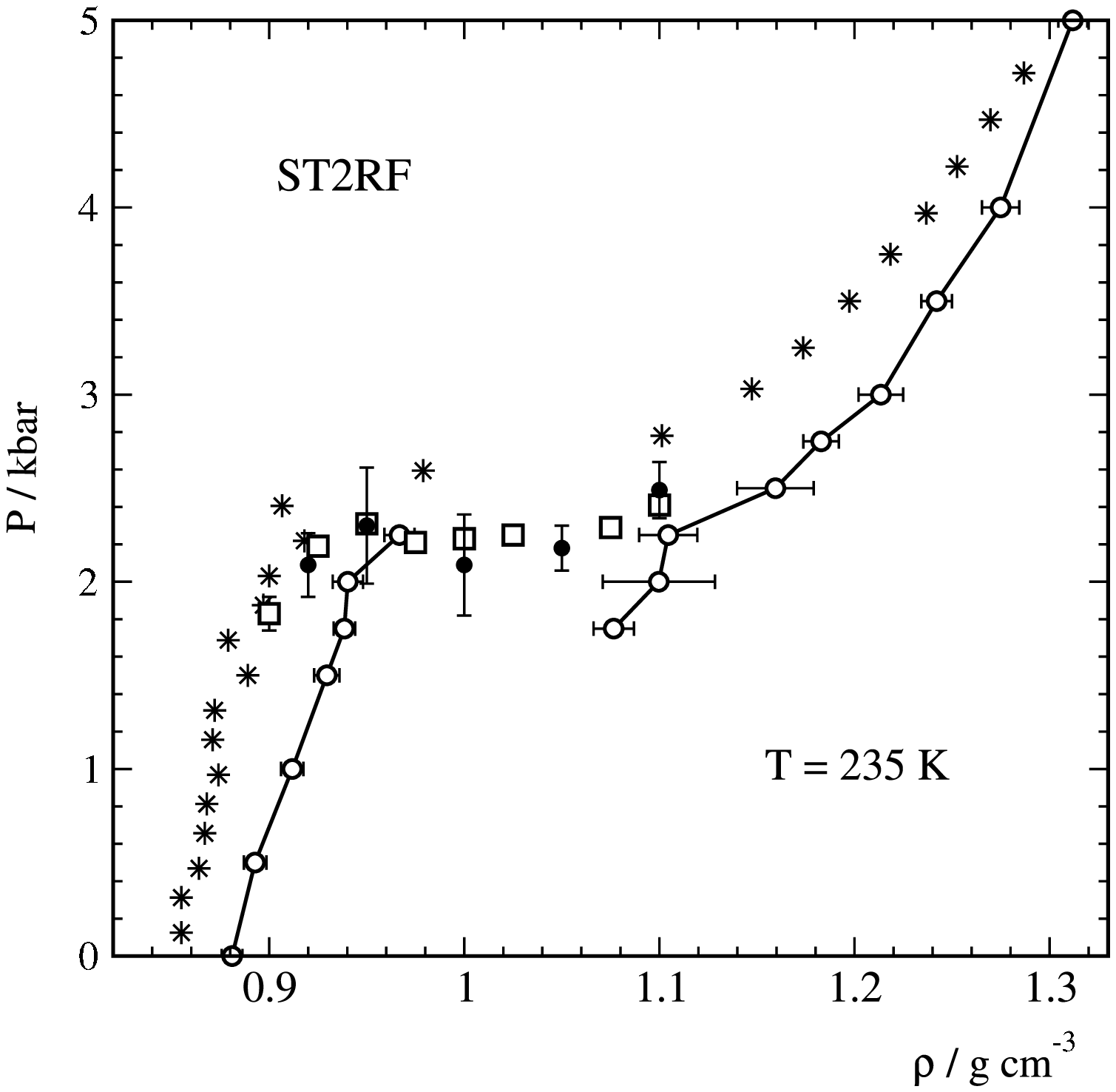}
\caption{ Isotherm of ST2RF water at \textit{T} = 235 K from MC simulations in the restricted \textit{NPT} ensemble (open circles) and from MD simulations in the \textit{NVT} ensemble (solid circles \cite{Poole93a}, squares \cite{Har97a}). The isotherm, obtained from a sedimentation profile at \textit{T} = 230 K \cite{Sciort03} is shown by stars.}
\end{figure} 
\begin{figure}[t]
\includegraphics[width=6cm]{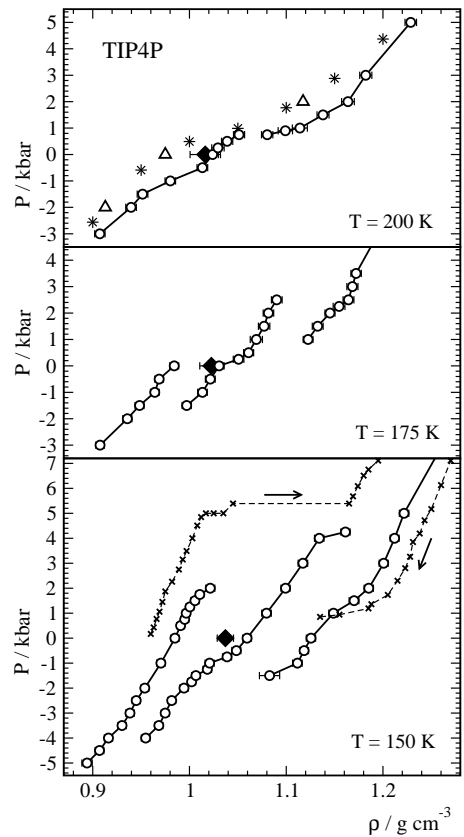}
\caption{ Isotherms of TIP4P water from MC simulations in the restricted \textit{NPT} ensemble (open circles) and densities of the liquid phase at the liquid-vapor coexistence curve (diamonds at \textit{P} = 0 bar). Data points from MD simulations in the \textit{NVT} ensemble \cite{Poole93a} are shown by stars and from MD simulations in the \textit{NPT} ensemble at \textit{T} = 193 K \cite{Tanaka96} are shown by triangles. Experimental data points, corresponding to the transition between low-density and high-density amorphous ice at \textit{T} = 110 K, \cite{Brazhkin1,Brazhkin2} are shown by crosses. Arrows indicate compression and decompression.}
\end{figure} 
\section{\label{sec:Discussion}Discussion}
Phase transitions between two liquid or amorphous phases with different densities were observed for several water models by simulations in the restricted \textit{NPT} ensemble. The reliability of the applied method was tested in various ways. Firstly by comparison with direct GEMC simulations of the liquid-vapor and liquid-liquid coexistence of ST2 water. In particular, the \textit{P} = 0 isobar, simulated in the restricted \textit{NPT} ensemble from the supercooled region up to \textit{T} = 450 K well agrees with the liquid branch of the liquid-vapor coexistence curve obtained by GEMC simulations (see Fig.2 in Ref.\onlinecite{BGO03}). Additionally, we have tested the sensitivity of the simulation results in the restricted \textit{NPT} ensemble to the number of molecules in the subcells. An increase of the system from 513 molecules (27 subcells, containing 19 molecules each) to 621 molecules (27 subcells, containing 23 molecules each) does not affect the results notably (see Fig.12, \textit{T} = 260 K). This increase of the number of molecules in the subcells is accompanied by an increase of the density fluctuations in the metastable states and a further increase of the number of molecules per subcell can lead to phase separation in the single subcells.  
The ability of the simulations in the restricted \textit{NPT} ensemble to locate correctly the phase transitions can be illustrated by using the triple (liquid-liquid-vapor) point of ST2 water as an example. The densities of the coexisting liquid phases, obtained at about zero pressure in the restricted \textit{NPT} ensemble are close to those obtained from the simulations of the liquid-vapor coexistence in the Gibbs ensemble and from the direct equilibration of two liquid phases in the Gibbs ensemble (see Fig.12 and also Fig.4 in Ref.\onlinecite{BGO03}). 
\par
There is a general agreement of our simulation results with available simulation studies of the same water models. Discrepancies are connected mainly with the differences in implementation of the water model. In particular, the use of long-range corrections for the intermolecular interaction in water models, which were parameterized without any long-range corrections, essentially affects the liquid density. The use of LRCI corrections causes a decrease of the liquid density especially at low temperatures. This effect is most dramatic for the ST2 water model and therefore we consider in fact two water models: ST2 and ST2RF. The use of LRLJ corrections results in the opposite effect, i.e. in an increase of the liquid density. Again, this effect enhances with decreasing temperature.
\begin{figure}[t]
\includegraphics[width=6cm]{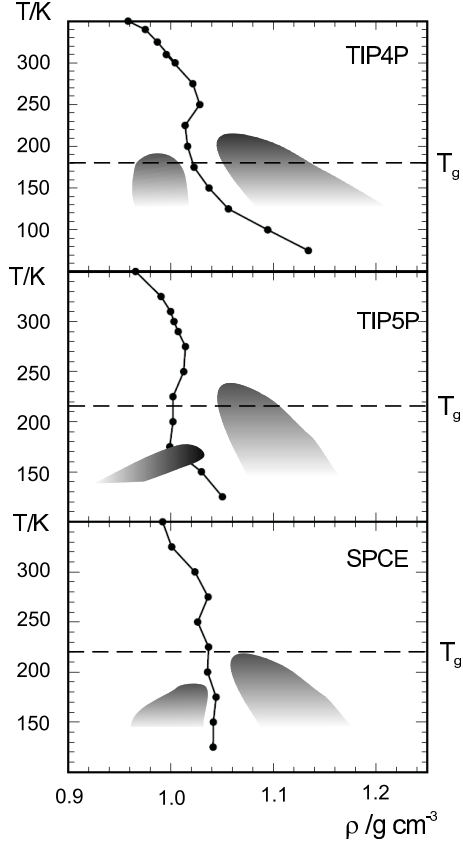}
\caption{ Schematic representation of the location of the liquid-liquid phase transitions (shadowed areas) with respect to the liquid branch of the liquid-vapor coexistence curve (solid circles) for TIP4P, TIP5P and SPCE water models. Glass temperature at \textit{P} $\approx$ 0 bar is indicated by dashed line.}
\end{figure}
\par
Another discrepancy between different simulation studies of supercooled water originates from the method, used for the detection (location) of the liquid-liquid phase transitions. Due to the intrinsic limitations of the simulations in the canonical \textit{NVT} ensemble (see Introduction), this method notably underestimates the critical temperature of the phase transition. For example, in some temperature interval below the critical temperature, where the simulations in the restricted ensemble evidence a phase transition, the isotherms, obtained in the \textit{NVT} ensemble, have a supercritical appearance (see Fig.16, \textit{T} = 200 K; Fig.18, \textit{T} = 225 K; Fig.19, \textit{T} = 200 K). Note also, that the applicability of simulations in the \textit{NVT} ensemble for the study of phase transitions is doubtful in the case of multiple phase transitions, which occur in a narrow pressure range (see, for example Fig.12, \textit{T} = 235 and 260 K), due to unavoidable and unpredictable distortions of the subcritical isotherms. 
\begin{figure}[t]
\includegraphics[width=6cm]{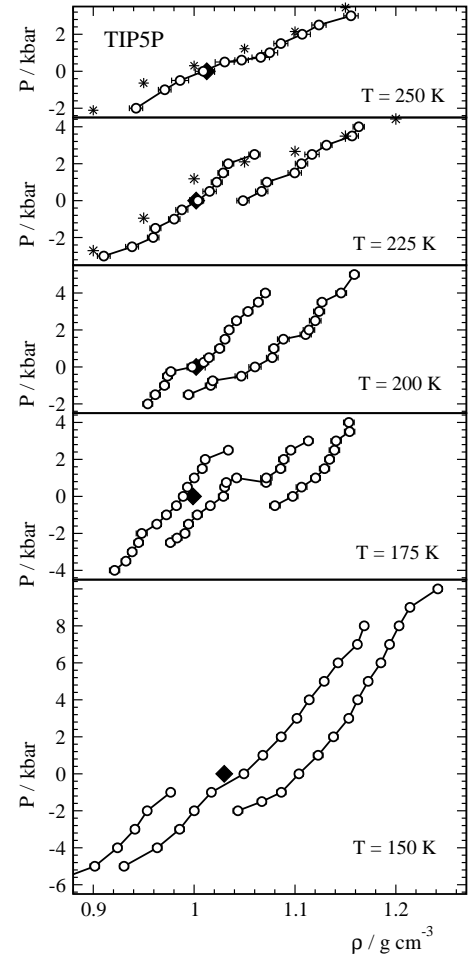}
\caption{Isotherms of TIP5P water from MC simulations in the restricted NPT ensemble (open circles) and densities of the liquid phase at the liquid-vapor coexistence curve (diamonds at \textit{P} = 0 bar). Data points from MD simulations in the \textit{NVT} ensemble \cite{Sciort02} are shown by stars (for \textit{T} = 225 K these points were obtained by interpolation of the data for \textit{T} = 220 and 230 K).}
\end{figure}       
\par
The phase diagrams of the liquid-vapor and the liquid-liquid phase transitions are highly sensitive  to the water model. All studied water models (except the ST2RF model) show comparable liquid densities at ambient conditions, where all these models were parameterized. Both upon heating and cooling, these models show increasing deviations from the saturated liquid density. For instance, the liquid-vapor critical temperature \textit{T$_c$} varies from $\sim$ 540 K to $\sim$ 640 K for the different water models (see Table I). The influence of the water model is even more obvious, when the distance between \textit{T$_c$} and the temperature \textit{T$_{max}$} of the liquid density maximum is considered (Table I). This distance is largest and comparable with the experimental value for the three-site SPCE water model, whereas its is much smaller for five-sites water models (ST2, ST2RF and TIP5P). Below \textit{T$_{max}$} and at supercooled temperatures the liquid density at the liquid-vapor coexistence curve and the densities of the coexisting liquid phases of the liquid-liquid phase transitions are highly sensitive to the water model and its implementation.  
\par
The multiplicity of liquid-liquid phase transitions, which we obtained for five non-polarizable models and Jedlovszky and Vallauri \cite{Jedl} obtained for a polarizable water model, seems to be more comprehensible, than the existence of only one liquid-liquid transition. Obviously, these transitions are related to the variety of amorphous phases of the one-component isotropic fluid, which differ in their local order. It is natural to assume, that each crystalline phase (or at least the crystalline phases, which can directly transform to the liquid phase) should result in a corresponding amorphous (liquid) phase with a short-range order, which is reminiscent of the crystalline phase.\cite{Tan00,Braz99} So, a multiplicity of liquid-liquid phase transitions may be expected for water, showing at least 12 crystalline phases, 5 of which are in direct equilibrium with the liquid,\cite{Petr}. This corroborates with the existence of at least 3 amorphous phases of supercooled water: low-density amorphous ice (LDA), high-density amorphous ice (HDA) \cite{Mishima85} and very-high-density amorphous ice (VHDA). \cite{Loert} Moreover, the existence of other amorphous phases of supercooled water can not be excluded.\cite{Brow,Tulk} 
\begin{figure}[t]
\includegraphics[width=6cm]{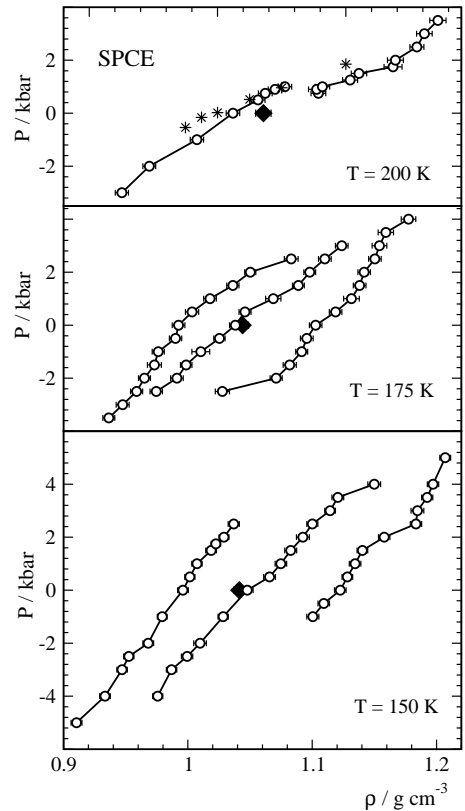}
\caption{ Isotherms of SPCE water from MC simulations in the restricted \textit{NPT} ensemble (open circles) and densities of the liquid phase at the liquid-vapor coexistence curve (diamonds at \textit{P} = 0 bar). Data points at \textit{T} = 200 K from MD simulations in the \textit{NVT} ensemble \cite{Har97} are shown by stars.  }
\end{figure}    
\par
The existence of more than one liquid-liquid phase transition in a one-component system was also concluded from simulation studies of other model fluids.\cite{Bul03,Deb01} An isotropic soft-core intermolecular potential with two steps causes the appearance of one liquid-liquid phase transition. \cite{Fr1,Fr2} Two liquid-liquid phase transitions were observed, when a third step was included in the soft-core potential.\cite{Bul03} Two liquid-liquid phase transitions were also observed for a waterlike model fluid confined between hydrophobic surfaces.\cite{Deb01}  
\par
We have found three liquid-liquid phase transitions of the ST2 water model. Hence, there are 4 distinct phases of supercooled water, which we have numbered as phase I to IV, starting from the phase with the lowest density. The density range, where the liquid-liquid transitions occur, is found to be the widest for ST2 and ST2RF water (this coincides with the wide hysteresis loop obtained for the ST2RF model by quick isothermal compression/decompression \cite{Poole93b}). The density interval of the two lower-density transitions in the ST2 model practically coincides with the density range of the first transition in the ST2RF model. Thus, 3 supercooled phases (phases I, III and IV) are found for the ST2RF model. Two liquid-liquid phase transitions and 3 supercooled phases (phases I and, presumably, phases II and III) we have found for TIP4P, TIP5P and SPCE water. The density range of the liquid-liquid transitions in each of these models is much narrower than in the ST2 and ST2RF models. The latter observation corroborates the narrow hysteresis loop obtained for the TIP4P model by quick isothermal compression/decompression.\cite{Poole93b}   
\par
The density of LDA (the lowest-density phase of amorphous water observed experimentally) is about 0.94 g cm$^{-3}$ at zero pressure and \textit{T} = 77 K.\cite{Mishima85} At the same conditions, the density of HDA was reported from 1.14-1.15 g cm$^{-3}$ in Ref.\onlinecite{Loert} to 1.17-1.19 g cm$^{-3}$ in Ref.\onlinecite{Mishima85}), whereas the density of VHDA is about 1.25-1.26 g cm$^{-3}$.\cite{Loert} Comparing with the densities of our isotherms at the same pressure we can attribute LDA to the lowest density phase (phase I) of the first liquid-liquid phase transition of all studied water models. HDA likely corresponds to phase III in all studied models, whereas VHDA could be attributed to phase IV, which is observed in ST2 and ST2RF water only. 
\par
Phase II, which is observed for all studied water models, except the ST2RF model, usually corresponds to liquid water in equilibrium with the vapor (see the diamonds in Figs.12,16,18 and 19): in the whole studied temperature range of the TIP4P and SPCE water models, above the triple point for the ST2 model and below the triple point for the TIP5P model. The density of phase II in the supercooled region of the ST2, SPCE and TIP5P models is about 1.00-1.04 g cm$^{-3}$ and varies slightly with the temperature.
For the TIP4P water model phase II achieves the density 1.14 g cm$^{-3}$ with decreasing temperature at zero pressure. The search of such a phase in real supercooled water seems to be a challenge for the experimental studies. There has been a single experimental observation of amorphous water (hyperquenched water) with the density 1.04 g cm$^{-3}$, reported in Ref.\onlinecite{Brow}. 
\par
Experimentally, transitions between various amorphous ices can be studied directly at temperatures below \textit{T} $\approx$ 150 K, i.e. below the temperature of the spontaneous crystallization of the amorphous water. At these temperatures the phase transition between HDA and LDA occurs at a pressure of \textit{P} $\approx$ 2 kbar, \cite{MishimaJCP,Brazhkin1,Brazhkin2} which slightly increases upon cooling.\cite{MishimaJCP} Measurements of the low-temperature metastable melting curves of crystalline ices \cite{MishimaPRL} provide an extension of the HDA/LDA transition line up to about 230 K and down to pressures of about +0.5 kbar. The densities of the coexisting phases at the HDA/LDA phase transition is about 0.94 and 1.20 g cm$^{-3}$ at \textit{T} = 130 K \cite{MishimaJCP} and they do not change essentially upon cooling to \textit{T} = 110 K.\cite{Brazhkin1,Brazhkin2} Experimental data points for the isotherm \textit{T} = 110 K,\cite{Brazhkin1,Brazhkin2} are compared with the simulated \textit{T} = 150 K isotherm of TIP4P water in Fig.16. The experimental hysteresis loop approximately covers the density interval, which corresponds to the two liquid-liquid (amorphous-amorphous) transitions of the TIP4P, TIP5P and SPCE water models. This confirms our assignment of phase I to LDA and phase III to HDA. A comparison of the experimental and simulated data, shown in Fig.16 (lower panel) may also be used as an indication for the existence of phase II in real water. The location of the phase transitions of the TIP4P model in the pressure-density plane  is closer to the experiment, than of TIP5P and SPCE models. This conclusion is in line with the results of computer simulations,\cite{Sanz,Vega} which show the ability of the TIP4P model to describe correctly the phase diagram of crystalline ices, unlike the SPCE and TIP5P models. Note finally, that the ST2 water model likely underestimates the densities of HDA and VHDA due to the overestimation of the tetrahedral water structure.     
\par
Other common (universal) features of the phase behavior can be noticed. In particular, the second transition in all studied models (except the specific case of the ST2RF model) seems to be rather universal. It is always located at positive pressures and its coexistence pressure increases with decreasing temperature. The proximity of the critical point of this transition to the liquid density maximum and to the consecutive density minimum suggests to attribute the existence of these features to the influence of the second liquid-liquid phase transition. Indeed, above \textit{T}$_{max}$ the influence of the liquid-vapor critical point on the density of the liquid branch is dominant and the temperature derivative of its density is negative, i.e. it shows normal behavior. Below \textit{T}$_{max}$ the derivative ($\partial\rho$/$\partial$T$)$$_P$ is positive and the liquid branch can be considered as a supercritical isobar (\textit{P} $\approx$ 0) of the second liquid-liquid phase transition. With further decreasing temperature, the zero pressure isobar turns to the "normal" behavior with ($\partial\rho$/$\partial$T$)$$_P$ $<$ 0 after the density minimum. This "normal" behavior of the isobar (\textit{P} $\approx$ 0) in deeply supercooled region can be attributed to the tilt of the two-phase region of the second liquid-liquid phase transition toward higher densities with decreasing temperature (Fig.17) rather than to the influence of the distant liquid-vapor critical point. So, the liquid density minimum and its consecutive maximum appear to be the result of a crossover between these three regimes. Note, that the critical temperature of the second liquid-liquid phase transition in all studied models appears above (or close to) the glass transition temperature. 
\par
The first liquid-liquid phase transition is much more sensitive to the choice of the water model and its implementation. This transition is located completely at positive pressures only for the ST2RF water model, which does not provide a realistic water density at ambient pressures. It is located completely at negative pressures (TIP4P and SPCE models), or it meets the liquid-vapor coexistence curve in a triple point (ST2 and TIP5P models). The critical point of this transition is located at slightly negative pressures for the ST2, TIP4P and SPCE models and at slightly positive pressure in the case of the TIP5P model. Only for the ST2 and TIP4P models the critical point of the first liquid-liquid phase transition is above the estimated zero pressure glass transition temperature. For the TIP5P and SPCE water models this transition has be considered as a transition between two glassy states. 
\par
The line of the first liquid-liquid phase transition in the \textit{P-T} plane is characterized by an essentially negative slope for the ST2 model and an essentially positive slope for the TIP5P model, whereas it could not be determined for the TIP4P and SPCE models. The presence of a triple point, where the vapor coexists with two liquid phases, means, that the liquid branch of the liquid-vapor coexistence curve crosses the liquid-liquid spinodal at some temperature below the triple point temperature. Such crossing can explain the apparent singular behavior of some thermodynamic properties, which in real liquid water is observed at $\approx$ 228 K, \cite{Speedy} i.e. 49$^\circ$ below the  temperature of the density maximum. This scenario is very close to that observed for the ST2 model, where the triple point is located about 35$^\circ$ below\textit{ T}$_{max}$ and the liquid-liquid spinodal crosses the liquid-vapor coexistence curve at about 40$^\circ$ below \textit{T}$_{max}$.\cite{BGO03} The triple point for the TIP5P model is found about 100-125$^\circ$ below \textit{T}$_{max}$ and therefore the liquid-liquid spinodal should cross the liquid-vapor coexistence curve far away from \textit{T}$_{max}$. 
\par
The variation of the heat capacity of liquid water with decreasing temperature seems to be highly sensitive to the location of the first liquid-liquid phase transition. In the TIP4P and SPCE water models, which do not show a triple point, noticeable temperature changes of \textit{C}$_p$ upon cooling can not be detected. A pronounced increase of \textit{C}$_p$ upon cooling is observed for the TIP5P, ST2 and ST2RF water models. The first two models show a triple point of the first liquid-liquid and the liquid-vapor phase transition.  
\par
The structure of liquid water in various thermodynamic states can be studied by the analysis of clustering and percolation.\cite{Rahman} In particular, the apparent singular behavior of some properties of liquid water in the supercooled region was attributed to the percolation of "four-bonded" water molecules.\cite{ST} This idea can be incorporated naturally in the observed phase behavior of supercooled water. Indeed, the percolation line of physical clusters of one of the constituting components should coincide with the spinodal of the phase transition, cross the coexistence curve in the critical point and extend into the one phase region.\cite{Coniglio,Klein} Thus, crossing the spinodal of a phase transition is equivalent to crossing a percolation line. The analysis of water clustering allowed the localization of the percolation line in aqueous solutions with respect to the liquid-liquid phase transition.\cite{percol2002} We have applied this method also to identify the molecular species with different local order, which may be responsible for the liquid-liquid phase transitions in supercooled ST2 water.\cite{BGO03,BGO05} The first, lowest-density liquid-liquid transition is caused by the formation of an infinite hydrogen-bonded network of perfectly coordinated tetrahedral water molecules. With decreasing temperature along the liquid-vapor coexistence curve, the spinodal of the first liquid-liquid phase transition is crossed, which is the percolation line of the perfectly coordinated tetrahedral water molecules (Fig.4 in Ref.\onlinecite{BGO03}).   
\par
Based on our simulations of the phase diagrams of various water models in the supercooled region, we can speculate about the possible location of liquid-liquid phase transitions in real water. We may expect, that there are at least two liquid-liquid phase transitions in real water, which effect the properties of liquid water at ambient conditions. One transition should be located at positive pressures with its critical point close to the density maximum of liquid water. This transition seems to be responsible for the anomalous behavior of the liquid water density. Another transition may meet the liquid-vapor transition and the spinodal of this transition could be the origin of the experimentally observed thermodynamic singularities of liquid water at 228 K.\cite{Speedy} 
\par
Finally, we would like to mention some perspectives for further studies of the phase behavior of fluids. The existence of various liquid phases of a one-component isotropic fluid is related presumably to different kinds of short and intermediate range order in the liquids. Indeed, the liquid phase with the lowest density consist mainly of perfectly tetrahedrally coordinated water molecules, whereas the liquid water phase with the highest density consists mainly of water molecules with close packed Lennard-Jones-like local order.\cite{BGO03} The main structural features of the other water phase(s) remain unclear and deserve further studies. The structure of these phases has to be analyzed in more detail, taking into account the various kinds of local ordering in crystalline ices.  
\par
In accord with the liquid-vapor phase transition, the fluid density is usually considered as the order parameter of the liquid-liquid phase transition of a one-component fluid. On the other hand, the concentration is the order parameter of liquid-liquid phase transitions in mixtures. In the case of a one-component fluid it seems to be reasonable to consider the concentration of molecules, showing some definite kind of local order, as an order parameter of the liquid-liquid phase transition. Such an approach is reminiscent of the well-known two- and multi-states water models (see, for example Ref.\onlinecite{Robinson}), originating from the ideas of H.Whiting and W.C.Roentgen of the 19th century.\cite{Whiting,Roentgen} Note, that the mutual transformations between different kinds of local ordering in a one-component fluid violates the conservation of the concentrations of species. Thus, finding a more appropriate order parameter of the liquid-liquid phase transition in a one-component fluid is an actual problem of the physics of liquids. Moreover, the universality class of this transition is not necessarily the universality class of the Ising model, as in the case of binary mixtures. For example, the possibility to change the local order continuously introduces unavoidable disorder, which could vary with the thermodynamic conditions. This means, that the critical behavior of such a system could belong to the universality class of random field Ising models. We can not exclude, that the enhancement of disorder with increasing temperature could lead to a rounding of the phase transition and the disappearance of a true critical point. In this case the two coexisting liquid phases are not infinite and the two-phase state appears as a domain structure.    
\begin{acknowledgments}
We thank Deutsche Forschungsgemeinschaft, Forschergruppe 436, for financial support.
\end{acknowledgments}
\newpage  

\pagebreak
\bibliography{mult}          
\end{document}